\documentclass[14pt,preprint2,longabstract]{aastex}
\newcounter{column_number}
\shorttitle{A {\it Chandra} survey of Galactic globular clusters}
\shortauthors{Cheng et al.}
\begin{document}
\title{Exploring the Mass Segregation Effect of X-ray Sources in Globular Clusters. III. Signs of Binary Disruption in M28}
\author{Zhongqun Cheng$^{1}$, Huijun Mu$^{2}$, Zhiyuan Li$^{3,4}$, Xiaojie Xu$^{3,4}$, Wei Wang$^{1}$ and Xiangdong Li$^{3,4}$}
\affil{$^{1}$ School of Physics and Technology, Wuhan University, Wuhan 430072, People's Republic of China}
\affil{$^{2}$ School of Physics and Microelectronics, Zhengzhou University, Zhengzhou 450001, People's Republic of China}
\affil{$^{3}$ School of Astronomy and Space Science, Nanjing University, Nanjing 210023, People's Republic of China}
\affil{$^{4}$ Key Laboratory of Modern Astronomy and Astrophysics (Nanjing University), Ministry of Education, Nanjing 210023, People's Republic of China}
\email{chengzq@whu.edu.cn}

\begin{abstract}
Using archival {\it Chandra} observations with a total effective exposure of 323 ks, we derive an updated catalog of point sources in the bulge globular cluster M28.
The catalog contains 502 X-ray sources within an area of $\sim475\, \rm arcmin^{2}$, and more than $90\%$ of these sources are first detected in this cluster.
We find significant dips in the radial distribution profiles of X-ray sources in M28, with the projected distance and width of the distribution dip for bright ($L_{X} \gtrsim 4.5\times 10^{30} {\rm\ erg\ \,s^{-1}}$) X-ray sources are larger than the faint ($L_{X} \lesssim 4.5\times 10^{30} {\rm\ erg\ \,s^{-1}}$) sources.
The ``generalized King model" fitting give a slightly larger average mass for the bright sources ($1.30\pm0.15\,M_{\odot}$) than the faint sources ($1.09\pm0.14\,M_{\odot}$), which support a universal mass segregation delay between heavy objects in GCs.
Compared with 47 Tuc and Terzan 5, we show that the dynamical age of M28 is comparable to Terzan 5 and much smaller than 47 Tuc, but it is evolving more fast (i.e., with smaller two-body relaxation timescale) than 47 Tuc. These features may suggest an acceleration effect of cluster dynamical evolution by tidal shock in M28.
Besides, we find an abnormal deficiency of X-ray sources in the central region ($R \lesssim 1.5 \arcmin$) of M28 than its outskirts, which indicate that M28 may have suffered an early phase of primordial binary disruption within its central region, and mass segregation effect will erase such a phenomenon as cluster evolve to older dynamical age.

\end{abstract}

\section{Introduction}

As one of the ancient self-gravitating systems in the Universe, the structure of globular clusters (GCs) is subtly balanced by the production of energy in the core and the outflow of energy from the system until their evolution to final core collapse \citep{henon1961}. The two-body relaxation is the fundamental process driving cluster evolution, as it dominates the transportation of energy and mass in GCs. Through two-body relaxation, stars are driven to reach a state of energy equipartition, lower-mass stars therefore tend to obtain energy and escape from cluster (thus leading to the energy outflow in GCs); whereas main-sequence (MS) binaries and massive stars tend to lose energy and sink in the gravitational potential well of GCs, where strong dynamical interactions between binaries and other stars could take place frequently. Depending on the binary hardness (defined as $\eta=|E_b|/E_{k}$, with $E_{b}$ the bound energy of binaries and $E_{k}$ the average stellar kinetic energy in GCs), soft binaries ($\eta << 1$) tend to absorb energy from fly-by stars and become softer or disrupted, whereas hard binaries ($\eta >> 1$) tend to transfer energy to fly-by stars and become harder (i.e., the binary burning processes; \citealp{hills1975,heggie1975}). Hard binaries therefore can strongly influence the evolution of GCs---sufficient to delay, halt or even reverse the core-collapse \citep{heggie2003,fregeau2003}.

Theoretically, there are many products that can be dynamically formed through the encounters of hard binaries in GCs, which include intermediate-mass black holes (IMBH; \citealp{miller2002,portegies2002,portegies2004,gurkan2004,giersz2015,fragione2018}), binary compact objects that can contribute to gravitational wave sources \citep{portegies2000,downing2010,samsing2014,rodriguez2015,rodriguez2016,askar2017,hong2018,ye2020}, low-mass X-ray binaries (LMXBs; \citealp{rasio2000,ivanova2010,giesler2018,kremer2018a}), millisecond pulsars (MSPs; \citealp{ivanova2008,ye2019}), cataclysmic variables (CVs; \citealp{ivanova2006,shara2006,belloni2016,hong2017,belloni2017,belloni2019}), coronally active binaries (ABs) and blue straggler stars (BSS; \citealp{fregeau2004,chatterjee2013}).
As GCs aged, it is natural to infer that more and more binary burning products be generated in the core of GCs. However, not all of these products could be retained by the cluster, since hard binaries may receive a large recoil velocity and be ejected from the host cluster  \citep{downing2011,morscher2013,morscher2015,bae2014,kremer2018b}.

At present, there is still no unambiguous evidence for IMBH in Galactic GCs \citep{tremou2018,mann2019,abbate2019,greene2019}, while 8 candidates of stellar mass black holes (BHs) have been found in several GCs \citep{strader2012,chomiuk2013,millerjonse2015,bahramian2017,shishkovsky2018,giesers2018,giesers2019}. For neutron star (NS) systems, about 21 NS-LMXBs \citep{vandenberg2019} and more than 130 MSPs have been identified in Galactic GCs\footnote{http://www.naic.edu/~pfreire/GCpsr.html}. For comparison, the abundances (i.e., number per unit stellar mass) of BH and NS systems are about $\sim100-1000$ times higher in GCs than in the Galactic field \citep{clark1975,katz1975,camilo2005,ransom2008,generozov2018}, which suggests that binary burning favors or assists creating these objects in GCs.
However, when extending these studies to CVs and ABs, \citet{cheng2018a} found that most GCs have a slightly lower X-ray emissivity (i.e., X-ray luminosity per unit stellar mass, which is a good proxy of CV and AB abundance in GCs) than the Galactic field, although the dynamical evolution of binaries is found to obey the Hills-Heggie law in GCs \citep{cheng2018b}.
These evidence suggest that, at least, the primordial binary formation channel of CVs and ABs is suppressed in most GCs.

A possible explanation is that the binary burning processes are taking place sequentially in GCs. Since the two-body relaxation timescale is anticorrelated with the mass of the heavy object, stars with different mass are expected to sink to the cluster center with different speeds (i.e., there is a mass segregation delay between different heavy objects in GCs).
Therefore, with a larger average mass than the normal stars, BH and NS are more likely to concentrate to cluster center and take part in the binary burning processes, which lead to an overabundance of exotic BH and NS systems in GCs than in the Galactic field. Whereas similar processes are inefficient for white dwarfs (WD), due to the much smaller average mass. Even worse, progenitor MS binaries of CVs are more massive than WDs, thus they are more likely to concentrate to cluster core and suffer strong dynamical encounters. Contrary to binary burning processes that transform MS binaries into CVs or ABs, simulations suggest that most of the primordial binaries will be dynamically disrupted in the core of GCs \citep{davies1997,belloni2019}.

The mass segregation delay of heavy objects has been confirmed in two massive GCs, 47 Tuc and Terzan 5.
Utilizing archival {\it Chandra} data, \citet{cheng2019a,cheng2019b} performed a deep survey of weak X-ray sources in 47 Tuc and Terzan 5. They found significant dips in the surface density distribution profile of X-ray sources in these two clusters, with the locations (i.e., projected distance from the GC centers) and widths of the distribution dips for bright sources being larger than that of the faint sources. The distribution dips are thought to be created by the mass segregation of heavy objects in GCs \citep{ferraro2012}, while their difference in locations and widths may represents a mass segregation delay between the considered objects.
Indeed, the average mass of the bright X-ray sources is estimated to be more massive than the faint sources, thus they drop to the cluster center faster and their distribution dips will propagate outward further \citep{cheng2019a,cheng2019b}. More interestingly, a comparison between these two clusters shows that the dynamical evolution of Terzan 5 is faster (i.e., with smaller two-body relaxation timescale), but its dynamical age is significantly younger than 47 Tuc. Since Terzan 5 is located much closer to the Galactic center than 47 Tuc, these features may suggest that tidal stripping effect is effective in accelerating the dynamical evolution (thus the binary burning processes) of GCs \citep{cheng2019b}.

In this work, we perform a similar study of X-ray sources mass segregation effect in M28, focusing on revealing the dynamical disruption of binary stars in the cluster core.
M28 is located in the inner Galactic region ($l=7.7982^{\circ}$,$b=-5.58068^{\circ}$), at a distance of $d=5.5$ kpc from the Sun \citep{harris1996}.
The orbit of M28 was found to be highly eccentric, with perigalactic distance of $R_{p}=0.57\pm0.1$ kpc, and apogalactic distance of $R_{p}=2.9\pm0.23$ kpc \citep{baumgardt2019}, which indicates that M28 may have suffered from strong tidal stripping in the Milky Way.
The structure of M28 was found to be relatively compact, with a half light radius of $R_{h}=1.97\arcmin$ (corresponding to $3.15\,\rm pc$) and a total mass of $\sim 3.7\times 10^{5}\,M_{\odot}$ \citep{harris1996}.
On the other hand, M 28 is found to host 12 MSPs \citep{bogdanov2011}. This is the third largest population of known pulsars in GCs, after that of Terzan 5 (38) and 47 Tuc $(25)^{1}$.
These features make M28 a valuable laboratory for studying stellar dynamical interactions and cluster dynamical evolution.

The remainder of this paper is organized as follows. In Section 2, we describe the X-ray data analysis and the catalog creation. We study the X-ray source radial distribution in Section 3, and explore its relation to mass segregation, Galactic tidal stripping and binary dynamical evolution in Section 4. A brief summary of our main conclusions is presented in Section 5.

\section{X-ray Data Analysis}
\subsection{{\it Chandra} Observations and Data Preparation}

So far, there are 8 {\it Chandra} observations, all taken with the Advanced CCD Imaging Spectrometer (ACIS), with the aimpoint located at the S3 CCD. From level 1 events file, we used the {\it Chandra} Interactive Analysis Observations (CIAO, version 4.11) and the Calibration Database (version 4.8.4) to reprocess the data, following the standard procedure\footnote{http://cxc.harvard.edu/ciao}. To inspect the possible background flares, we also created background light curve for each observation by excluding the source regions. A slight background flare was found in ObsID 2683, we therefore removed the episodes of background flares from this observation. The total effective exposure amounts to 323 ks in the central region of M28.
We searched for X-ray sources within the field of view (FoV) of each observation, and utilized the ACIS Extract (AE; \citealp{broos2010}) package to refine the source positions. Then, we corrected the relative astrometry among the X-ray observations. ObsID 9132, which has the longest exposure time, was adopted as the reference frame.
Finally, we combined the 8 observations into merged event files. Three groups of images have been created in soft (0.5--2 keV), hard (2--8 keV), and full (0.5--8 keV) bands and with a bin size of 0.5, 1, and 2 pixels, respectively.
As a demonstration, the merged full-band image (with binsize of 1 pixel) was illustrated in Figure-\ref{fig:smoothimage}.
Our data preparation details are summarized in Table 1.

\subsection{Sources Detection and Sensitivity}

Following the steps used in 47 Tuc and Terzan 5 \citep{cheng2019a,cheng2019b}, we employed a two-stage approach to create the X-ray sources catalog in M28. Firstly, we ran {\it wavdetect} on each of the 9 combined images, using the ``$\sqrt{2}$ sequence" wavelet scales and aggressive false-positive probability thresholds to find weak X-ray sources. If possible X-ray sources were missed by the {\it wavdetect} script, we also inspected the merged images visually and add them into the detection source lists. The {\it wavdetect} results were then combined into a master source list, which resulted in a candidate list of 576 sources. Owing to the relatively loose source-detection thresholds, this list was expected to enroll some false sources.
Therefore, we utilized the AE package to extract and filter the candidate source list. As stated in the AE handbook\footnote{http://personal.psu.edu/psb6/TARA/ae\_users\_guide.pdf}, AE was developed for analyzing multiple overlapping {\it Chandra} ACIS observations, which enable us to extract and evaluate the properties of X-ray sources interactively and iteratively.
AE provides an important parameter (i.e.,the binomial no-source probability $P_{B}$) to evaluate the significance of a source, which is defined on the null hypothesis that a source does not exist in the source extraction aperture, and the observed excess number of counts over background is purely due to background fluctuations \citep{weisskopf2007}. The formula to calculate $P_{B}$ is given by
\begin{equation}
P_{\rm B}(X\ge S)=\sum_{X=S}^{N}\frac{N!}{X!(N-X)!}p^X(1-p)^{N-X},
\end{equation}
where $S$ and $B$ is the total number of counts in the source and background extraction region, $N=S+B$ and $p=1/({\rm 1+BACKSCAL})$, with $\rm BACKSCAL$ being the area ratio of the background and source extraction regions. A larger $P_{\rm B}$ value indicates that a source has a larger probability of being false.

\begin{figure*}[htp]
\centering
\includegraphics[angle=0,origin=br,height=0.5\textheight, width=0.9\textwidth]{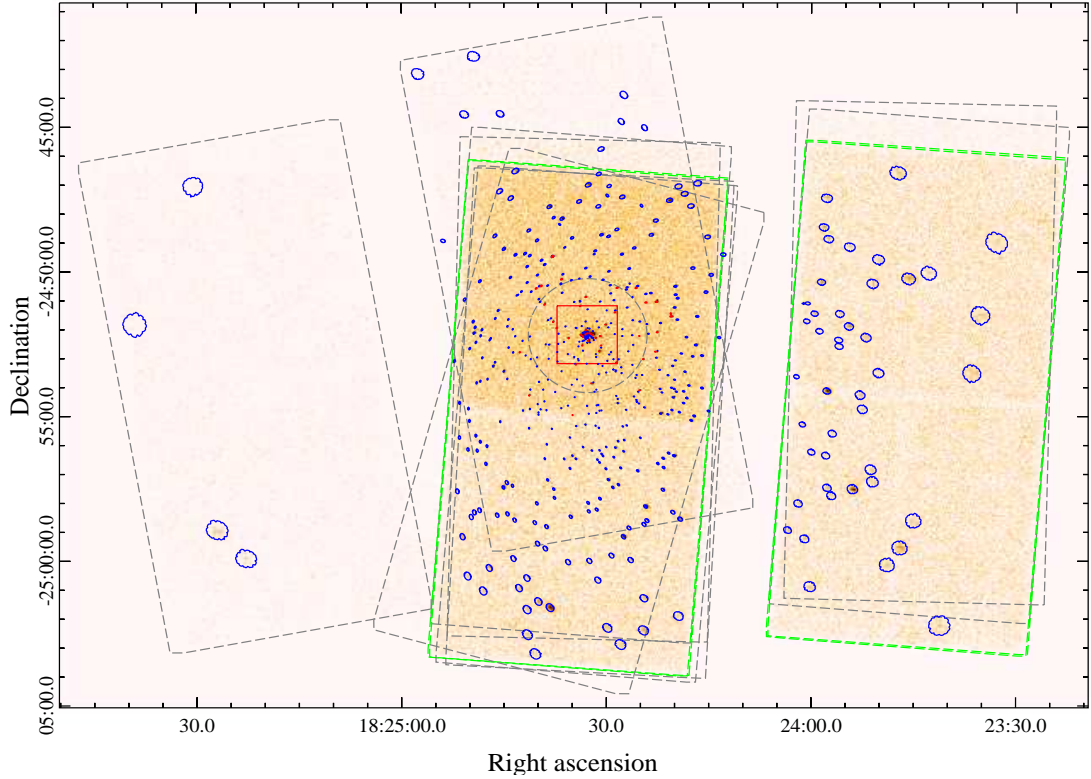}
\linespread{0.7}
\caption{The full-band (0.5-8 keV) {\it Chandra} merged image of M28. The images are smoothed with a Gaussian kernel with a radius of 3 pixels. Fields of view of the 8 observations in Table-1 are shown as dashed boxes. Sources detected by \citet{becker2003} are marked with red ellipses, while new detections of this work are shown as blue ellipses. Size of the ellipses representing the $\sim 90\%$ enclosed PSF power. Due to the much larger variation in detection sensitivity, only X-ray sources located within the field of view of ObsID 9132 and ObsID 9133 (green boxes) are used to calculate the surface density profile in Figure-\ref{fig:surfd}. A zoom-in view of the central $2\arcmin\times 2\arcmin$ region (red square region) is illustrated in the left panel of Figure-2. Grey dashed circle marks the half-light radius ($R_{h}=1.97\arcmin$) of M28. \label{fig:smoothimage}}
\end{figure*}

\begin{figure*}[htp]
\centering
\includegraphics[angle=0,origin=br,height=0.4\textheight, width=0.51\textwidth]{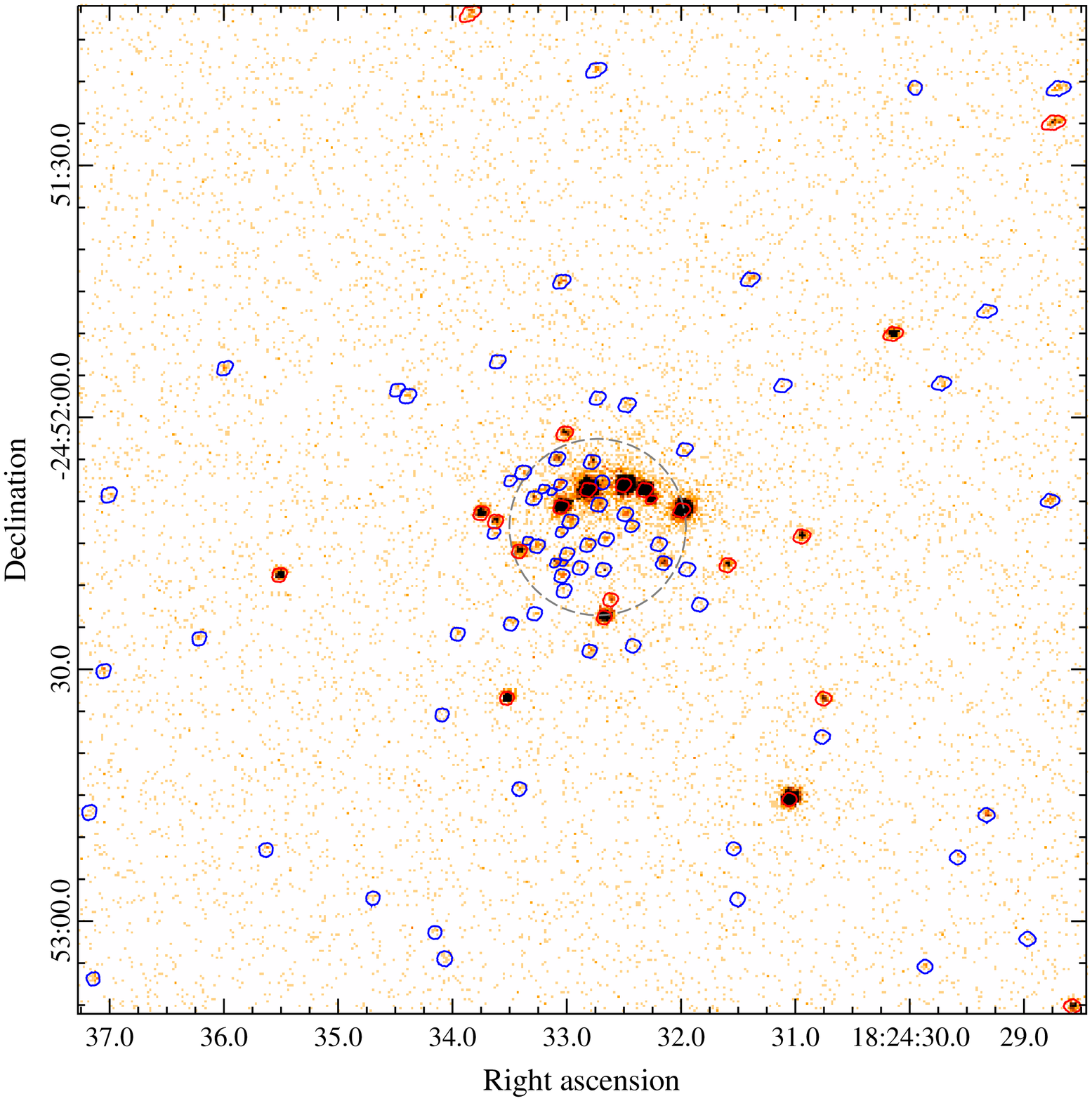}\includegraphics[angle=0,origin=br,height=0.4\textheight, width=0.53\textwidth]{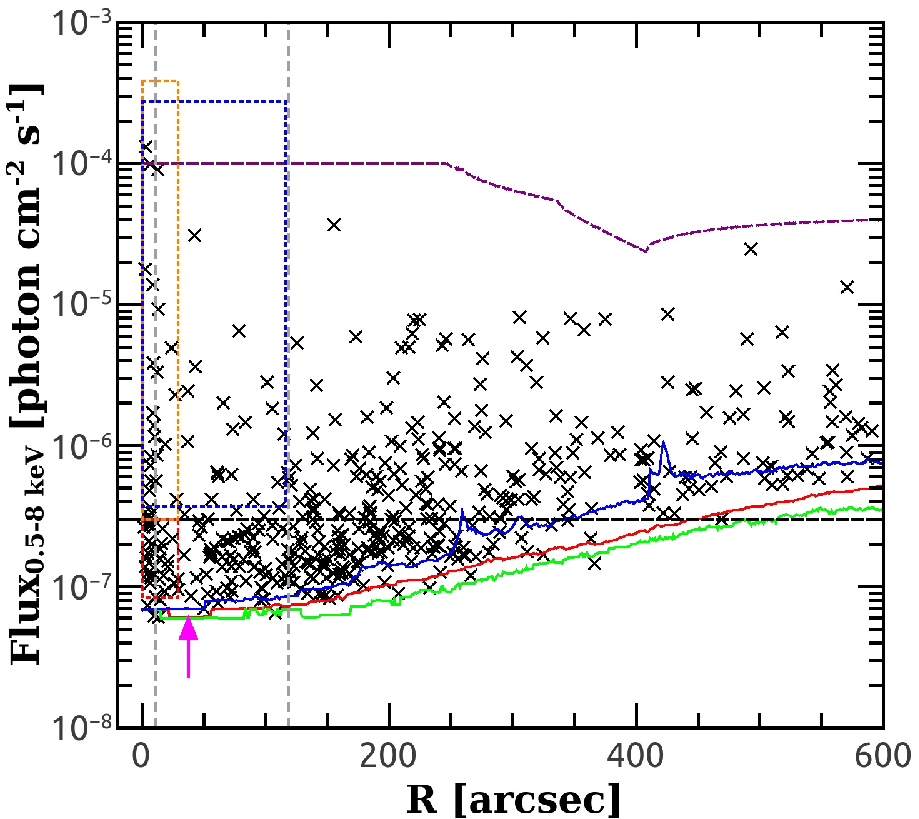}
\linespread{0.7}
\caption{Left: {\it Chandra} merged image of the central $2\arcmin \times 2\arcmin$ region of M28. The image was binned to $0\farcs25$ pixel$^{-1}$. The color-coded symbols denote the different types of sources, as in Figure-\ref{fig:smoothimage}. The grey dashed circle represents the cluster core radius ($R_{c}=10\farcs5$; \citealp{miocchi2013}). Right: 0.5-8 keV photon flux as a function of projected distance from the cluster center. The color-coded solid curves represent the median (red), minimum (green), and maximum (blue) limiting detection count rate at corresponding radial bins, while the purple dashed curve donates the coverage fraction ($\times 10^{-4}$) of the source analyzing region (i.e., green dashed boxes in Figure-\ref{fig:smoothimage}), respectively. The cluster core and half-light radii are displayed as grey dashed vertical lines, and the position of the distribution dip is marked with the magenta arrow. With a threshold photon flux of $F_{p}=3.0\times 10^{-7}$ $\rm photon\, cm^{-2}\, s^{-1}$ (black dashed horizontal line), we divided the X-ray sources into the faint and bright groups and analyze their surface density profiles in Figure-\ref{fig:surfd}. Samples selected by the dotted boxes are shown in Figure-\ref{fig:fit}, which have the advantage of unbiased detection (see Section 4 for details). \label{fig:roomin}}
\end{figure*}

During our AE usage, we set a more stringent threshold value of $P_{\rm B}>0.001$ for invalid sources. As experienced in 47 Tuc, this threshold $P_{\rm B}$ value was proved to be helpful in optimizing the source completeness and reliability.
Then, by extracting, merging, pruning, and repositioning the candidate sources interactively and iteratively, we obtained a stable list that all remaining sources meet the $P_{\rm B}$ condition in either full, soft, and hard bands.
Our final source catalog contains 502 X-ray sources. They are marked by ellipses in Figure-\ref{fig:smoothimage} and the left panel of Figure-\ref{fig:roomin}.
As a comparison, we also highlight the X-ray sources identified by \citet{becker2003} as red ellipses in the figures. Due to the much shallower exposure (i.e., $\sim 37$ ks from ObsID 2683, 2684 and 2685) and smaller source searching region ($R\leq 3.1\arcmin$), only 46 X-ray sources were detected in their work.

As suggested in Figure-\ref{fig:smoothimage}, the merged effective exposure maps of M28 are found to have a large variation within the {\it Chandra} FoV, which may lead to significant variation in detection sensitivity across the source analyzing region. Following the procedures employed in 47 Tuc \citep{cheng2019a}, we utilized the merged event images to create the detection sensitivity maps. Briefly, we first masked out the source regions from the merged event files, and refilled them with the surrounding pixels. The source-free counts images were utilized to generate the background maps. Then, at each pixel of the survey field, we modeled its source and background areas with local point-spread function (PSF; given by the exposure time weighted averaged PSF maps) and $\rm BACKSCAL$ values (given by nearby sources from AE), they are used to estimate the background counts (B). With given $\rm BACKSCAL$, B and threshold $P_{B}$ value, we can derive a limiting source counts (S) by solving Equation (1). Finally, we computed for each pixel the limiting detection count rates with the exposure map, and then converted them into limiting fluxes by assuming an absorption power-law model with photon index $\Gamma=1.4$ and column density $N_{\rm H}=2.32\times10^{21}\,\rm cm^{-2}$. Here $N_{\rm H}$ is derived from the color excess $E(B-V)$ of
M28 \citep{cheng2018a}.

\subsection{Catalog of Sources}

For the 502 sources in the catalog, we extracted their X-ray photometry and spectra with AE. The default source aperture was defined to enclose a PSF power fraction of $\sim 90\%$, which was allowed to reduce to a minimum value of $\sim40\%$ for very crowded sources. For the background regions, we built them with the AE ``BETTER\_BACKGROUNDS" algorithm, and their scales were set to enclose at least a minimum number of 100 background counts through the AE ``ADJUST\_BACKSCAL" stage.
We computed the net source counts in soft, hard and full band, respectively. Among the 502 X-ray sources, we found about one-third (i.e., $177/502$) of them have net counts greater than $\sim 30$ in the full band. Using the AE automated spectral fitting script, we modeled their spectra with an absorbed power-law spectrum, with the photon index was set as free parameters and the neutral hydrogen column density ($N_{\rm H}$) was constrained to no less than $N_{\rm H}=2.32\times10^{21}\,\rm cm^{-2}$.
For sources with net counts less than $\sim 30$, we converted their net count rates into unabsorbed energy fluxes using a power-law model. During the conversion, we fixed the photon index at $\Gamma=1.4$ and froze the absorption column density at $N_{\rm H}=2.32\times10^{21}\,\rm cm^{-2}$, and adopted the AE-generated merged spectral response files to calibrate the flux estimation.

Finally, we collated the source extraction and spectral fitting results into a main X-ray source catalog, with source labels sorted by their R.A.. Assuming a distance of 5.5 kpc for M28 \citep{harris1996}, we calculated for each X-ray source an unabsorbed luminosity in the soft, hard and full bands, respectively. By adopting an optical central coordinate $\alpha =18^{h}24^{m}32^{s}.73$ and $\delta =-24\arcdeg 52\arcmin 13.07\arcsec$ for M28 \citep{miocchi2013}, we also computed for each source a projected distance from the cluster center. The final catalog of point sources is presented in Table 2.

\section{Analysis: Source Radial Distribution}

Before the analysis of X-ray source surface density distribution in M28, we first check the source detection sensitivity across the merged {\it Chandra} FoV. Due to the large difference of exposure time between ObsID 9132/9133 and other observations (Table 1), the variation in limiting detection sensitivity is found to be over one order of magnitude. 
To minimize the uncertainty, we choose the FoV of ObsID 9132/9133 (green boxes in Figure-\ref{fig:smoothimage}) as the X-ray source radial distribution analyzing region in this paper, which has a deep and relatively flat detection sensitivity.
The limiting detection sensitivity of the ObsID 9132/9133 FoV is illustrated as a function of cluster projected radius in the right panel of Figure-\ref{fig:roomin}, with the red, green, and blue solid line represent the median, minimum, and maximum detection limit at each radial bin, respectively.
For comparison, we also displayed the photon flux of each source sample as black cross in the figure. It's evident that almost all of the X-ray sources are located above the minimum sensitivity line.
Due to the constraint of the ObsID 9132/9133 FoV, the coverage factor $f$ (defined as $dA/2\pi r dr$, with $2\pi r dr$ being the area of cluster radial bins and $dA$ the corresponding coverage area of the ObsID 9132/9133 FoV.) is less than 1 at larger cluster radii. We multiplied $f$ with $10^{-4}$ and plot it as purple dashed line in the figure.

\begin{figure*}[htp]
\centering
\includegraphics[angle=0,origin=br,height=0.4\textheight, width=1.05\textwidth]{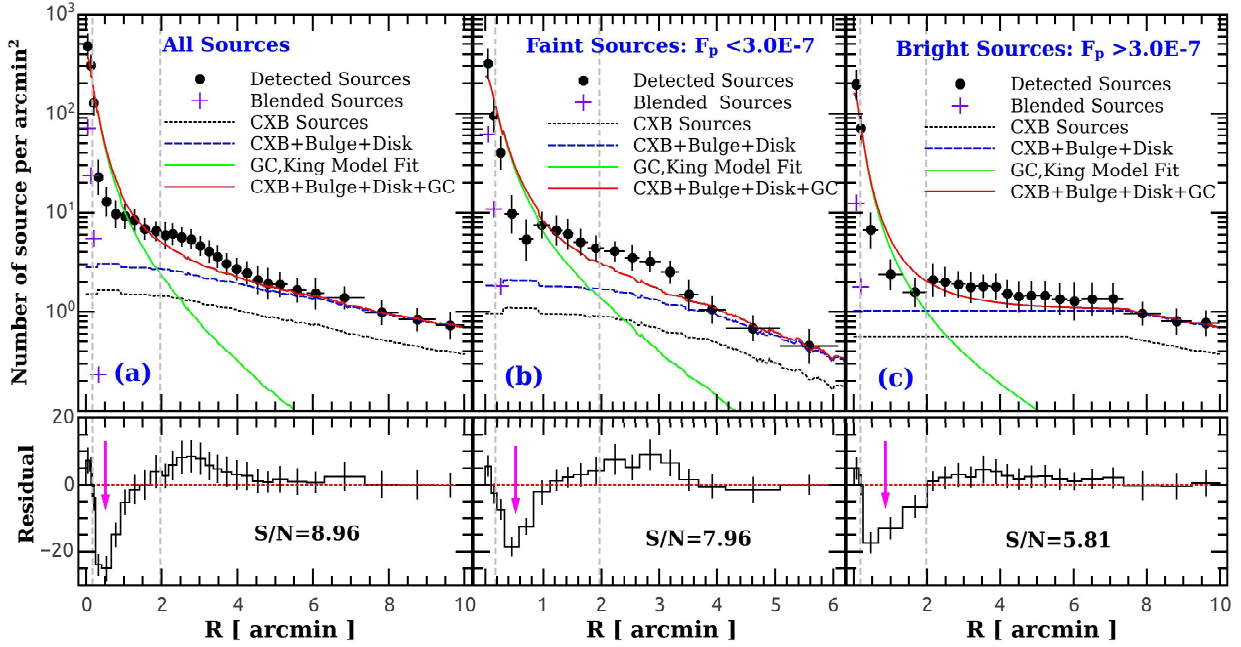}
\linespread{0.7}
\caption{Radial surface density distribution of X-ray sources in M28. The full band (0.5-8 keV) sample of X-ray sources are displayed as black dots in (a), we divided them into the faint (b) and bright (c) groups. The contributions of CXB sources are shown as black dotted curves, while the blue dashed curves represent the sum of CXB, Galactic Bulge and Disk X-ray sources components. The green solid lines mark the profile of stellar density convolving with the limiting sensitivity function, which was calculated from the best-fitting King model of M28 and have been normalized to match the total number of GC X-ray sources in each group. The red solid lines represent the sum of CXB ($N_{\rm CXB}$), Galactic Bulge and Disk ($N_{\rm G}$), and the King model components ($N_{\rm K}$).
The lower panels show the residual of the observed X-ray sources ($N_{\rm X}+N_{\rm B}$) subtracting the model predicted X-ray sources ($N_{\rm CXB}+N_{\rm G}+N_{\rm K}$). The black text represents the maximum significance of the distribution dips. The cluster core and half-light radii are displayed as grey dashed vertical lines.
X-ray sources are slightly overabundant in cluster center and scarce at the distribution dips (marked with magenta arrows), and then significantly exceed the predictions of the cluster light at larger radii. \label{fig:surfd}}
\end{figure*}

A total number of 467 X-ray sources are selected within the ObsID 9132/9133 FoV (also with $R<10\arcmin$), we plot their surface density profile as black dots in Figure-\ref{fig:surfd}(a).
Compared with a King model profile that describes the radial surface density distribution of normal stars in GCs, we found the distribution profile of X-ray sources in M28 decreases rapidly outside of the cluster core radius, then reach a plateau at $1\arcmin \lesssim R\lesssim 3\arcmin$, and there is a distribution dip around $R\sim 0.6\arcmin$ (see also the source distribution in the right panel of Figure-\ref{fig:surfd}).
According to \citet{cheng2019a,cheng2019b}, such a distribution dip could be caused by mass segregation of X-ray sources in GCs.
Since the mass segregation effect is found to be luminosity dependent in 47 Tuc and Terzan 5, we also divided the source sample into two subgroups according to their X-ray luminosities. To balance the source counts between the two groups of samples, we set a threshold photon flux of $F_{p}=3.0\times 10^{-7}$ $\rm photon\, cm^{-2}\, s^{-1}$ (or luminosity of $L_{X}\sim 4.5\times 10^{33}\, \rm erg\, s^{-1}$) and obtained 220 and 247 sources for the faint and bright groups of X-ray sources separately. The surface density profile of the faint and bright groups of X-ray sources is plotted as black dots in Figure-\ref{fig:surfd}(b) and Figure-\ref{fig:surfd}(c), respectively.
Again, significant distribution dips also can be found for these two subgroups of X-ray sources.

Due to source blending, the source density is more or less underestimated in the cluster center.
Assuming that a faint source could be easily blended by nearby brighter sources when it is located within the source extraction regions of the bright sources, we also corrected for blending effect in the central region of M28. For each concentric annulus with area of $A_{ann}$, the number of blended sources can be roughly estimated with function
\begin{equation}
N_{B}(>F_{p})=N_{X}(>F_{p})\beta,
\end{equation}
where $N_{X}(>F_{p})$ is the number of bright X-ray sources (with photon flux greater than $F_{p}$) located within $A_{ann}$, $\beta=A_{tot}/A_{ann}$, with $A_{tot}$ being the total source extraction regions of the bright X-ray sources.
The distribution profiles of the blended sources are presented as purple pluses in Figure-\ref{fig:surfd}. We estimated about 4 GC sources are blended in M28, with about 3 (1) of them belonging to the faint (bright) group of X-ray sources.

To identify the possible signal of mass segregation of X-ray sources in M28, we modeled the X-ray source distribution profiles with several components.
Following the procedures employed in 47 Tuc and Terzan 5 \citep{cheng2019a,cheng2019b}, we adopted the ${\rm log}N-{\rm log}S$ relations determined by \citet{kim2007} to estimate the contribution of the cosmic X-ray background (CXB) sources in the FoV.
The cumulative number count of CXB sources above a limiting sensitivity flux $S$ can be estimated with function
\begin{equation}
{N_{\rm CXB}(>S)}=2433(S/10^{-15})^{-0.64}-186\ {\rm deg^{-2}}.
\end{equation}
This equation is derived from Equation (5) of \citet{kim2007}, by assuming a photon index of $\Gamma=1.4$ for the {\it Chandra} ACIS observations in the 0.5-8 keV band.
With the limiting sensitivity maps obtained in Section 2.2, we calculated the CXB contributions and plot them as black dotted lines in Figure-\ref{fig:surfd}.

Since M28 is located very close to the Galactic center, the contamination of Galactic bulge and disk X-ray sources is non-negligible, as is the case in Terzan 5. Such a contamination also can be noticed in Figure-\ref{fig:surfd}, where the observed X-ray sources significantly exceed the predicted CXB components at large cluster radius. According to \citet{miocchi2013}, the surface density distribution of normal stars in M28 is dominant over the Galactic background stars within $R\sim 7\arcmin$. Therefore, we can estimate the Galactic background components by looking at the distribution profiles of X-ray sources outside of $R\sim 7\arcmin$. Assuming an uniform spatial distribution for the Galactic bulge and disk X-ray sources within the {\it Chandra} FoV, we calculated the Galactic background components ($N_{\rm G}$) with the limiting sensitivity map obtained in Section 2.2, which were normalized to match the surface density profiles of X-ray sources beyond $R\sim 7\arcmin$\footnote{Due to the decreases of detection sensitivity at larger cluster radii (see also the left panel of Figure-\ref{fig:roomin}), no sources were detected beyond $R\sim 7\arcmin$ for the faint group of X-ray sources, thus the Galactic component was normalized to match the difference value between the bright and the total source samples.}.
The sum of the CXB and the Galactic background components are shown as blue dashed lines in Figure-\ref{fig:surfd}. Then, the excess of X-ray sources over the blue dashed lines can be reasonably attributed to the GC X-ray sources (i.e., $N_{\rm X}+N_{\rm B}-N_{\rm CXB}-N_{\rm G}$).
Among the 467 X-ray sources, we found $\sim 215$ of them are from M28, and with about 125 (90) belonging to the faint (bright) group.

To better compare the radial distribution of X-ray sources with normal stars in M28, we plot the best fitting King model profiles of M28 as green solid lines in Figure-\ref{fig:surfd}. The best King model profiles were calculated with the cluster concentration ($c=2.01$) and the core radius ($R_{c}=10\farcs5$; \citealp{miocchi2013}), we normalized them to match the number of GC X-ray sources for each group of samples.
The sum of the CXB component ($N_{\rm CXB}$), Galactic background component ($N_{\rm G}$) and King model component ($N_{\rm K}$) are shown as red solid lines in the figure.
From Figure-\ref{fig:surfd}, it can be observed that there are significant differences between the distribution profiles of observed X-ray sources ($N_{\rm X}$) and model predicted sources in M 28. We calculated their residuals with equation $N_{\rm X}+N_{\rm B}-N_{\rm CXB}-N_{\rm G}-N_{\rm K}$, which are illustrated as a function of $R$ in the lower panels of Figures-\ref{fig:surfd}.
As showed in the figure, there is slight excess of X-ray sources than the model predictions in the core region of M28, which decreases to a minimum value at the distribution dip and then increases again at larger cluster radii.
Following the methods used in Terzan 5 \citep{cheng2019b}, we estimated the significance of the distribution dips with function ${\rm S/N}=(N_{\rm CXB}+N_{\rm G}+N_{\rm K}-N_{\rm X}-N_{\rm B})/\sqrt{N^{2}_{\rm CXB}\sigma^{2}_{c}+N^{2}_{\rm G}\sigma^{2}_{\rm G}+N^{2}_{\rm K}\sigma^{2}_{\rm K}+N^{2}_{\rm X}\sigma^{2}_{P}(1+\beta^{2})}$.
By adopting a nominal fitting error of $\sigma_{\rm G}=\sigma_{\rm K}=5\%$ for the King model fitting and Galactic background estimation, a CXB variance of $\sigma_{c}=17\%$ for the Chandra sources analyzing region in M28 \citep{moretti2009}, and the Poisson uncertainties $\sigma_{P}$ estimated with the formulae of \citet{gehrels1986} for detected X-ray sources, we adjusted the ranges of the distribution dips and estimated for each group of X-ray sources a maximum significance. The parameters of the distribution dips are summarized in Table 3, and their maximum significance values are marked as black text in the lower panels of Figure-\ref{fig:surfd}.
We found the median locations and widths (i.e., ranges of $N_{\rm X}+N_{\rm B}<N_{\rm CXB}+N_{\rm G}+N_{\rm K}$) of the distribution dips for the total, faint and bright groups of X-ray sources are $R_{dip}\simeq 45\arcsec$, $\Delta R_{dip}\simeq 75\arcsec$, $R_{dip}\simeq 35\arcsec$, $\Delta R_{dip}\simeq 60\arcsec$ and $R_{dip}\simeq 65\arcsec$, $\Delta R_{dip}\simeq 110\arcsec$, respectively.

Beside the distribution dips, it can be seen from Figure-\ref{fig:surfd} that there are significant distribution bumps of X-ray source in the outskirts of M28.
Contrary to the traditional idea that the GC X-ray sources are concentrated and overabundant in cluster center, these features may suggest an overabundance of X-ray sources in the outskirts of M28.
To quantify this phenomenon, we divided the survey area into several concentric annuli, and defined in each annulus the relative abundance ratio of X-ray sources as
\begin{equation}
R_{\rm X}=(N_{\rm X}+N_{\rm B}-N_{\rm CXB}-N_{\rm G})/{N_{\rm K}}.
\end{equation}
$R_{\rm X}>1$ ($R_{\rm X}<1$) indicate that there is a relative overabundance (deficiency) of X-ray sources in the annulus region\footnote{Note that $(N_{\rm X}+N_{\rm B}-N_{\rm CXB}-N_{\rm G})/N_{\rm K}=1$ when the full survey area (i.e., $0\arcsec \leq R \leq 600\arcsec$) is considered.}.
In Table-3, we calculated $R_{\rm X}$ within several radial bins in M28. It is evident that the relative abundance of X-ray sources is $\sim 150\%$ times larger than the prediction of normal stars within the cluster core. However, this excess (i.e., $N_{\rm X}+N_{\rm B}-N_{\rm CXB}-N_{\rm G}-N_{\rm K} \simeq 11$ within $R\leq 10\arcsec$) can not compensate the amount of sources missed in the distribution dip (i.e., with $N_{\rm X}+N_{\rm B}-N_{\rm CXB}-N_{\rm G}-N_{\rm K} \simeq -74$ within $10\arcsec \leq R\leq 90\arcsec$), and there is an abnormal deficiency of X-ray sources in the central region (i.e., $R_{\rm X} \sim 60\%$ within $R\lesssim 90\arcsec$) of M28  than its outskirts (i.e., $R_{\rm X} \sim 200\%$ at $R\gtrsim 90\arcsec$).

\section{Discussion}

\subsection{Mass Segregation of X-ray Sources in M28}
\begin{figure*}[htp]
\centering
\includegraphics[angle=0,origin=br,height=0.4\textheight, width=1.05\textwidth]{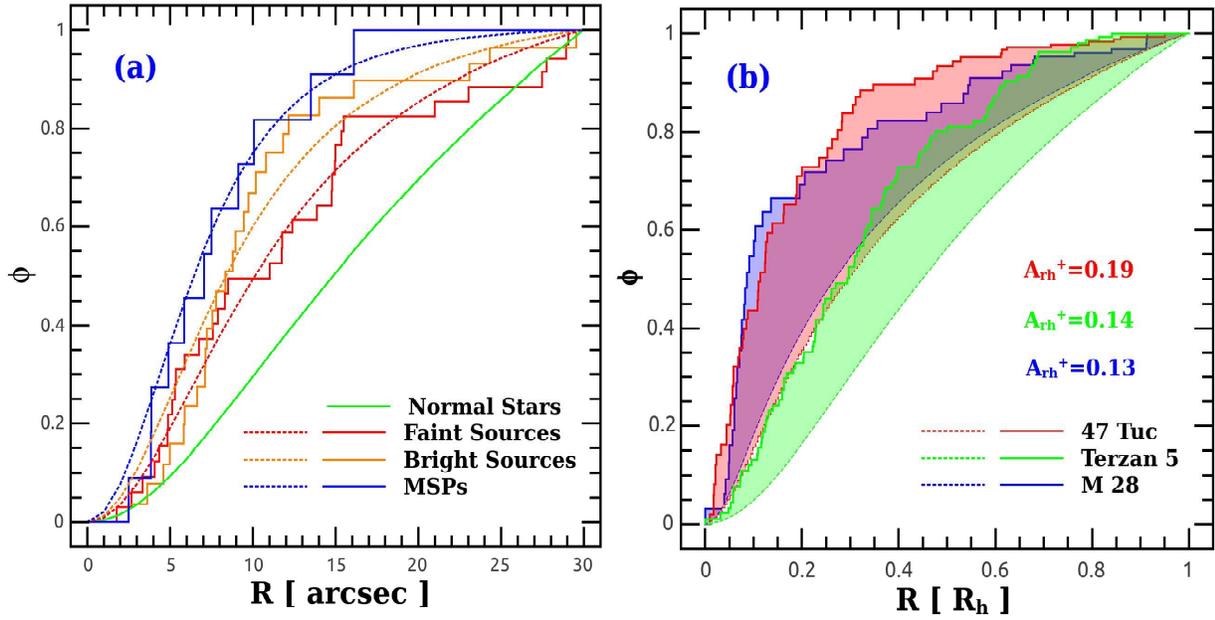}
\linespread{0.7}
\caption{(a): Generalized King model fitting of heavy objects in M28. The cumulative radial distribution of MSPs, bright and faint groups of X-ray sources (selected by the orange and red dotted boxes in Figure-\ref{fig:roomin}) are shown as blue, orange and red step lines, respectively. While the green solid line represents the King model predicted distribution of normal stars. The best-fitting results are shown as color-coded dotted lines. (b): Estimation of the dynamical age in M28 (blue), Terzan 5 (green) and 47 Tuc (red). The X-ray source sample of M28 was selected from the blue dotted box in Figure-\ref{fig:roomin}, while data for Terzan 5 and 47 Tuc were adopted from \citet{cheng2019b}. All X-ray sources are constraint to have a minimum unabsorbed luminosity of $L_{\rm X} \gtrsim 5.0\times 10^{30}\, \rm erg\,s^{-1}$ in the three GCs, we plot their cumulative distributions as solid step lines. For the distributions of the reference normal stars, they are calculated with the King model and are displayed as dotted lines. The area of the shaded regions ($A_{rh}^{+}$) is calculated with Equation (6), which is a good indicator of cluster dynamical age.  \label{fig:fit}}
\end{figure*}

According to Figure-\ref{fig:surfd}, the locations and widths of the distribution dips for bright X-ray sources are clearly larger than the faint X-ray sources, which may suggest a mass segregation delay between these two groups of X-ray sources in M28.
Similar to the procedures outlined in 47 Tuc and Terzan 5 \citep{cheng2019a,cheng2019b}, we estimated for each group of X-ray sources an average mass by fitting the surface density distribution with the ``generalized King model". This model assumes that all stars are dynamically relaxed in the cluster, thus the radial distribution of more massive objects (such as X-ray sources) will be more centrally concentrated than that of the reference normal stars.
To ensure that the selected sources are detection sensitivity unbiased and dynamically relaxed in M28, we set a minimum threshold photon flux of $8\times 10^{-8}$ $\rm photon\, cm^{-2}\, s^{-1}$ for the faint group of X-ray sources and constraint the source analyzing region as $R\leq 30\arcsec$ (i.e., source samples selected by the red and orange dotted boxes in the right panel of Figure-\ref{fig:roomin}). We then corrected background contamination and blending effect for each group of X-ray sources and plot their cumulative distributions as solid step lines in Figure-\ref{fig:fit}(a).
According to the generalized King model, the projected surface density profile of heavy objects can be expressed as
\begin{equation}
S(R)=S_{0}{\left[ 1+\left( \frac{R}{R_{c}}\right) ^{2}\right]}^{(1-3q)/2}.
\end{equation}
Here, $S_{0}$ is the normalization constant, and $q=M/M_{\ast}$ is the average mass ratio of heavy objects over the reference normal stars. $R_{c}=10.5\arcsec$ is the cluster core radius of M28, which was determined by stars over the red giant branch/subgiant branch/turn-off point or the upper main sequence \citep{miocchi2013}.

For comparison, we also illustrated cumulative radial distribution of MSPs as blue step line in Figure-\ref{fig:fit}(a). Among all the considered objects, the MSPs show the highest degree of central concentration, which was followed by the bright, faint groups of X-ray sources, and the reference normal stars. Using a maximum-likelihood method, we fit the cumulative radial distributions of the heavy objects with Equation (5), which yields a mass ratio of $q=2.38\pm0.35$, $q=1.85\pm0.22$ and $q=1.55\pm0.20$ for MSPs, the bright and faint X-ray sources, respectively\footnote{Here the errors are quoted at the 1$\sigma$ level.}. Assuming that the dominant visible stellar population has a mass of $M_{\ast}\sim 0.7\,M_{\odot}$ in M28 \citep{becker2003}, we derived an average mass of $1.67\pm 0.25 \, M_{\odot}$, $1.30\pm 0.15 \, M_{\odot}$ and $1.09\pm 0.14 \, M_{\odot}$ for MSPs, the bright and faint group of X-ray sources, respectively.
The best ``generalized King model" fitting results are shown as dotted lines in Figure-\ref{fig:fit}(a). The average mass of the bright group of X-ray sources is slightly larger than the faint X-ray sources, which is consistent with the observed distribution dips in M28.
In other words, the bright X-ray sources are more massive and their two-body relaxation timescale is shorter, thus they are expected to drop to the cluster center faster and their distribution dip will propagate outward further in M28.
We found that the sedimentation delay of X-ray sources in M28 is consistent with that found in 47 Tuc and Terzan 5 \citep{cheng2019a,cheng2019b}, supporting an universal mass segregation effect for X-ray sources in GCs.

\subsection{Acceleration of GC Dynamical Evolution by Tidal Stripping}

As suggested by \citet{ferraro2012}, the mass segregation effect of heavy objects can be used to build a ``clock" to measure the dynamical evolution age of GCs, and GCs with larger projected radius of distribution dips (i.e., $R_{dip}$ in unit of cluster core radius) are thought to be dynamically older.
In Figure-\ref{fig:surfd}, we found the projected radius of the distribution dip for bright and faint group of X-ray sources is $R_{dip}\simeq 60\arcsec\simeq 5.7R_{c}$ and $R_{dip}\simeq 35\arcsec\simeq 3.3R_{c}$ separately.
At first sight, values of $R_{dip}$ in M28 are smaller than that detected in 47 Tuc and Terzan 5 (i.e., with $R_{dip}\simeq 8.2R_{c}$ and $R_{dip}\simeq 7.8R_{c}$ for the bright sources, $R_{dip}\simeq 4.8R_{c}$ and $R_{dip}\simeq 3.9R_{c}$ for the faint sources, respectively; \citealp{cheng2019b}), which suggest that M28 is dynamically younger than Terzan 5 and 47 Tuc.
However, precision of $R_{dip}$ is subject to the binning of the heavy objects, which may hamper the comparison of dynamical age between GCs.
Therefore, we also estimated the dynamical age of M28 with the $A^{+}$ indicator, which is defined as the enclosed area between the cumulative radial distribution of heavy objects (here the X-ray sources $\phi_{\rm X}(R)$) and that of a reference stars ($\phi_{\rm REF}(R)$):
\begin{equation}
A^{+}(R)=\int_{R_{0}}^{R}  \phi_{\rm X}(R)-\phi_{\rm REF}(R)dR.
\end{equation}
As stated by \citet{alessandrini2016}, the $A^{+}$ indicator is binning-independent and more precise than $R_{dip}$, and it can be applied to all GCs \citep{lanzoni2016,ferraro2018}.

In Figure-\ref{fig:fit}(b), we assess the dynamical age of M28 with the $A^{+}_{rh}$ indicator, by integrating the value of $A^{+}$ from the cluster center to the half-light radius. For comparison, the data of Terzan 5 and 47 Tuc are also illustrated in the figure, they are adopted from \citet{cheng2019b}.
Considering that the mass segregation effect is luminosity dependent in M28, we first set a minimum threshold unabsorbed luminosity of $L_{\rm X} \gtrsim 5.0\times 10^{30}\, \rm erg\,s^{-1}$ for the X-ray source sample (i.e., same as in Terzan 5 and 47 Tuc). Then, we corrected the source blending and background contamination for the source sample, and plot their cumulative radial distribution as blue solid step line in Figure-\ref{fig:fit}(b). As in Terzan 5 and 47 Tuc, we adopted the King model to calculate the distribution of reference normal stars in M28, which is plotted as blue dotted line in the figure.
The integration yields a dynamical age of $A^{+}_{rh}=0.13$ for M28, which is comparable to Terzan 5 ($A^{+}_{rh}=0.14$) but much smaller than 47 Tuc ($A^{+}_{rh}=0.19$). Again, the smallest value of $A^{+}_{rh}$ indicates that M28 is dynamically the youngest among the three GCs.

However, the above suggestion is challenged by the other dynamical age indicator. With a smaller core relaxation timescle in M28 (${\rm log}(t_{relax})=7.62$) than in 47 Tuc (i.e., ${\rm log}(t_{relax})=7.84$; \citealp{harris1996}), M28 is expected to evolve more fast and it should be dynamically older than 47 Tuc. Indeed, considering that the age of M28 ($\tau\sim 12.5-13.0\, \rm Gyr$; \citealp{kerber2018}) is comparable to 47 Tuc ($\tau\sim 13.06\, \rm Gyr$; \citealp{forbes2010}), we can infer that the dynamical age of M28 ($\tau/t_{relax}\sim 300$) is larger than 47 Tuc ($\tau/t_{relax}\sim 190$). Obviously, these results are in conflict with what $R_{dip}$ and $A_{rh}^{+}$ suggest.

The possible explanation is that M28 may have experienced stronger tidal stripping in Milky Way, as discussed in the case of Terzan 5 \citep{cheng2019b}. The orbit of M28 is short and highly eccentric, with perigalactic distance $R_{p}=0.57\pm0.1$ kpc and apogalactic distance $R_{p}=2.9\pm0.23$ kpc, which is much closer to the Galactic center than 47 Tuc (with $R_{p}=5.46\pm0.01$ kpc $R_{p}=7.44\pm0.0.02$; \citealp{baumgardt2019}).
Accordingly, M28 will suffer from more stronger tidal shock in the Galactic field, which accelerates the evaporation rate of stars (thus increase the rate of energy outflow) from the cluster.
In response, the cluster will contract homologically, to increase the stellar density and enhance the energy production rate of the cluster core.
In fact, although the total mass of M28 ($M_{c}\sim 3.7\times 10^{5} M_{\odot}$) is about three times small than 47 Tuc ($M_{c}\sim 1.2\times 10^{6} M_{\odot}$; \citealp{cheng2018a}), we found the structure of M28 is more compact than 47 Tuc\footnote{The core, half-light and tidal radius of M28 is $R_{c}\simeq 0.28\,\rm pc$, $R_{h}\simeq 3.15\,\rm pc$ and $R_{t}\simeq 17.96\,\rm pc$ \citep{harris1996}. As a comparison, the corresponding size of 47 Tuc is $R_{c}\simeq 0.41\,\rm pc$, $R_{h}\simeq 3.71\,\rm pc$ and $R_{t}\simeq 41.56\,\rm pc$, respectively \citep{mcLaughlin2006}.}, with cluster central luminosity density ($\rho_{c}=10^{4.86}\, L_{\odot}\,\rm pc^{-3}$) is comparable to 47 Tuc ($\rho_{c}=10^{4.88}\, L_{\odot}\,\rm pc^{-3}$; \citealp{harris1996}), and specific stellar encounter rate ($\gamma=1750^{+226}_{-245}$) clearly larger than 47 Tuc ($\gamma=844^{+130}_{-113}$; \citealp{cheng2018a}).
Taking these aspects together, we argue that M28 is very similar to Terzan 5, they both have suffered from strong tidal stripping process in the Galactic field, and the tidal shock is efficient in accelerating their dynamical evolution.

\subsection{Mass Segregation and Disruption of Binary Stars in GCs}

In Section 3. we show that there is an abnormal deficiency of X-ray sources in the central region of M28 than its outskirts (Table-3). Such a phenomenon is challenging to our current knowledge of X-ray source distribution in GCs, which holds that the X-ray sources are concentrated in the cluster center, and their abundance is greater in the central region than in the outskirts of GCs.
According to \citet{cheng2019a}, there are two formation channels for X-ray sources in GCs. The first is the dynamical channel, which is dominant in the cluster core and may be responsible for the formation of exotic objects (i.e., LMXBs, MSPs, etc.) and bright CVs in GCs \citep{pooley2003,pooley2006}. Another one is the primordial binary channel, which is prevailing in the outskirts of GCs and is a significant contributor to the weak X-ray source (i.e., faint CVs and ABs) population in GCs \citep{cheng2018a,cheng2019a}. Compared with the dynamical channel, simulations suggest that the primordial binary channel is suppressed in cluster center, since most of the primordial binaries will be dynamically disrupted before they otherwise could evolve into weak X-ray sources in the dense core of GCs \citep{davies1997,cheng2018a,belloni2019}.
If this was the case, the abnormal deficiency of X-ray sources in the central region of M28 may represent a relic of early primordial binary disruption in the center of GCs.

In fact, we also found in literatures that there are similar peculiar radial distributions of MS binaries in star clusters. For example, by searching the F-star binary systems in the compact, $15-30\, \rm Myr$ old Large Magellanic Cloud cluster NGC 1818, \citet{degrijs2013} demonstrated a decreasing profile of MS binary fraction toward the center of NGC 1801. They argued that many soft MS binaries have been dynamically disrupted in the central region of this cluster, due to the much higher stellar encounter frequency. Whereas similar soft MS binaries may have survived from the dynamical disruption, provided that they are located in the outskirts of the star cluster \citep{degrijs2013}. However, \citet{geller2013,geller2015} argued that such a peculiar distribution profile of MS binaries seems to be short-lived in clusters, since mass segregation effect tends to erase these features as cluster evolve into older dynamical age.

Nevertheless, recent simulations of GCs have shown that more massive objects (such as stellar-mass BHs) may play a fundamental role in driving cluster evolution and shaping their present-day structure \citep{chatterjee2017,arcasedda2018,askar2018,weatherford2018,kremer2019a}.
Under the effect of mass segregation, the BHs are expected to concentrate to cluster center and form a high density subcluster\footnote{Traditionally, the BH subcluster was thought to be decoupled with the rest of the cluster and will lead to the ejection of most BHs on a few Gyr timescale \citep{kulkarni1993,sigurdsson1993}. However, modern state-of-the-art simulations of GCs indicate that the BH subcluster does not stay decoupled from the rest of GC for prolonged periods, since interactions that eject BHs from the subcluster also cause it to expand and re-couple with the host cluster, which dramatically increasing the timescale for BH evaporation \citep{breen2013,morscher2013,morscher2015,wang2016,kremer2019a}.} \citep{spitzer1969}, which may influence the dynamical evolution of primordial binaries in two ways.
First, the BH subcluster may quench or slow down the mass segregation of binaries in GCs \citep{weatherford2018}, and compared to clusters with few BHs, GCs with a large number of BHs are found to have a large core radius and low central density (thus longer two-body relaxation timescale; \citealp{merritt2004,hurley2007,chatterjee2017,arcasedda2018,askar2018}). Second, the BH subcluster may serve as a dominant internal energy source and suppress any other binary burning processes in GCs. For example, simulations show that the binary burning processes of NS are suppressed by BHs in GCs, and the number of dynamically formed MSPs is anti-correlated with the population of BHs retained in the host cluster \citep{ye2019}. Therefore, primordial binaries will either be disrupted or be transformed (i.e., through exchange encounters) into BH-MS binaries\footnote{Due to the low duty cycles for the active state of a mass-transferring BH \citep{kalogera2004}, BH-MS systems are hard to find via X-ray and radio emission, thus their contribution to GC X-ray source population is small.} when they sink into the BH subcluster, and their evolution (both dynamical and primordial channels) to CVs or ABs will be suppressed in the central region of GCs.
Taking these aspects together, we argued that it is possible to have a long-lived abnormal deficiency of weak X-ray sources in cluster central region than its outskirts.

Although the BH subcluster may play an important role in creating and maintaining the peculiar radial distribution of X-ray sources in M28, we argue that the population of retained BHs in this cluster is small. This is because not only the large population of MSPs, but also the apparent mass segregation effect of X-ray sources that found in this cluster. Indeed, we found in Section 3 that there are slight excess of X-ray sources in the core region of M28, which suggest that M28 may have run out of its BH population, thus NS and low-mass binaries started to drop into the cluster core and dominate the binary burning processes. If this was the case, mass segregation effect will gradually erase the deficiency of X-ray sources in the central region of M28. In fact, with a larger dynamical age than M28, we found the deficiency of X-ray sources in the central region of Terzan 5 is not as serious as M28 \citep{cheng2019b}. While for 47 Tuc, its dynamical age is much older than M28 and Terzan 5, most of the X-ray sources are concentrated in the cluster center and their relative abundance ratio in core region is much larger than that in the cluster outskirts \citep{cheng2019a}.

\section{Summary}

We have presented a sensitive study of weak X-ray sources in M28. By analyzing the radial properties of X-ray sources in this cluster, our main findings are as follows:

1. We detected 502 faint X-ray sources within an area of $\sim475\, \rm arcmin^{2}$ in M28. The cleaned net exposure time of the study is 232 ks, and more than $90\%$ of the sources are first detected in this cluster.

2. We found significant distribution dips in the surface density distribution profiles of X-ray sources in M28. The location and width of the distribution dip for the total, faint ($L_{\rm X}\lesssim 4.5\times 10^{30}\,\rm erg\, s^{-1}$) and bright ($L_{\rm X}\gtrsim 4.5\times 10^{30}\,\rm erg\, s^{-1}$) source samples is $R\sim 45\arcsec$, $\Delta R\sim 75\arcsec$, $R\sim 35\arcsec$, $\Delta R\sim 60\arcsec$ and $R\sim 65\arcsec$, $\Delta R\sim 110\arcsec$, respectively. And there is a delay of mass segregation between the faint and bright groups of X-ray sources in M28.

3. The ``generalized King model" fitting yields a larger average mass for the bright group of X-ray sources ($1.30\pm 0.15\, M_{\odot}$) than the faint sources ($1.09\pm 0.14\, M_{\odot}$), which is qualitatively consistent with the observed distribution dips in M 28, and suggest that mass segregation may be responsible for the creation of these distribution dips.

4. Compared with 47 Tuc and Terzan 5, we demonstrated that the dynamical age of M28 is comparable to Terzan 5 and much smaller than 47 Tuc, but it is evolving more fast (i.e., with samller two-body relaxation timescale) than 47 Tuc. Since M28 is located more close to Galactic center than 47 Tuc, these features may suggest a acceleration effect of cluster dynamical evolution by tidal shock in M28.

5. We show that there is an abnormal deficiency of X-ray sources in the central region ($R\lesssim 1.5\arcmin$) of M28 than its outskirts, which could be a relic of primordial binary disruption in the central region of GCs. Such a peculiar distribution profile will be erased by mass segregation effect as cluster evolve into older dynamcial age.

\begin{deluxetable}{lccrrrcrc}
\tablecolumns{8}
\linespread{1}
\tablewidth{0pc}
\tablenum{1}
\tablecaption{Log of {\it Chandra} Observations}
\tablehead{
\colhead{ObsID} & \colhead{Date} & \colhead{Exposure} & \colhead{Ra} & \colhead{Dec} & \colhead{Roll}  & \colhead{$\delta x$} & \colhead{$\delta y$} \\
\colhead{} & \colhead{} & \colhead{(ks)} & \colhead{(\arcdeg)} & \colhead{(\arcdeg)} & \colhead{(\arcdeg)} & \colhead{(pixel)} & \colhead{(pixel)}\\
\colhead{(1)} & \colhead{(2)} & \colhead{(3)} & \colhead{(4)} & \colhead{(5)} & \colhead{(6)} & \colhead{(7)} & \colhead{(8)}  }
\startdata
2683  & 2002 Sep 9  & 10.87  & 276.1327 & -24.8904 & 273.16 &  0.045 & -0.254  \\
2684  & 2002 Jul 4  & 12.74  & 276.1376 & -24.8900 & 285.35 &  0.284 & -0.252  \\
2685  & 2002 Aug 4  & 13.51  & 276.1333 & -24.8904 & 274.68 &  0.381 & -0.268  \\
9132  & 2008 Aug 7  & 142.25 & 276.1416 & -24.8904 & 274.46 &  0.000 &  0.000  \\
9133  & 2008 Aug 10 & 54.45  & 276.1416 & -24.8904 & 274.31 & -0.041 & -0.296  \\
16748 & 2015 May 30 & 29.65  & 276.1342 & -24.8651 & 79.66  &  0.338 &  0.231  \\
16749 & 2015 Aug 7  & 29.55  & 276.1412 & -24.8738 & 274.53 & -0.093 & -1.044  \\
16750 & 2015 Aug 11 & 29.56  & 276.1410 & -24.8740 & 271.41 &  1.271 & -0.341  \\
\enddata
\vspace{-0.5cm}
\tablecomments{Columns.\ (1) -- (3): {\it Chandra} observation ID, date and effective exposure time. Columns.\ (4) -- (6): telescope optical pointing positions and roll angle of each observation. Columns.\ (8) and (9): right ascension and decl. shift.}
\label{tab:obslog}
\end{deluxetable}

\begin{deluxetable}{cllccccccccc}
\tabletypesize{\scriptsize}
\tablewidth{0pt}
\tablecaption{Main {\it Chandra} Source Catalog}
\tablecolumns{12}
\linespread{1}
\tablewidth{0pc}
\tablenum{2}
\tablehead{
\colhead{XID} & \colhead{RA} & \colhead{Dec} & \colhead{Error} & \colhead{R} & \colhead{$\log P_{\rm B}$} & \colhead{$C_{net,f}$} & \colhead{$C_{net,h}$} & \colhead{$F_{p,f}$} & \colhead{$F_{p,h}$} & \colhead{$\log L_{X,f}$} & \colhead{$\log L_{X,h}$} \\
\colhead{} & \colhead{(\arcdeg)} & \colhead{(\arcdeg)} & \colhead{(\arcsec)} & \colhead{(\arcsec)} & \colhead{} & \colhead{(cts)}  & \colhead{(cts)}  & \colhead{(ph/s/cm$^{2}$)} & \colhead{(ph/s/cm$^{2}$)} & \colhead{(erg/s)} & \colhead{(erg/s)} \\
\colhead{(1)} & \colhead{(2)} & \colhead{(3)} & \colhead{(4)} &\colhead{(5)} & \colhead{(6)} & \colhead{(7)} & \colhead{(8)} & \colhead{(9)} & \colhead{(10)} & \colhead{(11)} & \colhead{(12)}}
\startdata
1 & 275.886925 & -24.816634 & 0.6 & 837.5 & $<$-5 & $108^{+134}_{-83}$  & $38^{+57}_{-20}$    & 2.52E-6 & 9.94E-7 & 31.58 & 31.42 \\ 
2 & 275.896753 & -24.858506 & 0.5 & 783.8 & $<$-5 & $154^{+177}_{-130}$ & $67^{+84}_{-50}$    & 3.49E-6 & 1.71E-6 & 31.69 & 31.55 \\ 
3 & 275.901918 & -24.892048 & 0.5 & 769.7 & $<$-5 & $207^{+231}_{-183}$ & $79^{+96}_{-62}$    & 4.91E-6 & 2.11E-6 & 31.80 & 31.62 \\ 
4 & 275.921525 & -24.037327 & 0.6 & 923.8 & $<$-5 & $225^{+249}_{-201}$ & $78^{+96}_{-61}$    & 6.98E-6 & 2.71E-6 & 32.11 & 31.76 \\ 
5 & 275.928294 & -24.834271 & 0.6 & 692.0 & $<$-5 & $76^{+93}_{-59 }$   & $10^{+19}_{-1}$     & 1.88E-6 & 2.74E-7 & 31.46 & 31.30 \\ 
6 & 275.937572 & -24.977005 & 0.5 & 754.2 & $<$-5 & $138^{+158}_{-119}$ & $59^{+74}_{-45}$    & 3.12E-6 & 1.49E-6 & 31.61 & 31.43 \\ 
7 & 275.940597 & -24.837490 & 0.3 & 650.3 & $<$-5 & $364^{+387}_{-341}$ & $201^{+219}_{-184}$ & 7.99E-6 & 4.89E-6 & 32.15 & 32.04 \\ 
8 & 275.945923 & -24.992580 & 0.3 & 761.8 & $<$-5 & $461^{+486}_{-435}$ & $207^{+225}_{-189}$ & 1.21E-5 & 6.05E-6 & 32.25 & 32.09 \\ 
\enddata
\vspace{-0.5cm}
\tablecomments{
Column\ (1): sequence number of the X-ray source catalog, sorted by R.A.
Columns.\ (2) and (3): right ascension and decl. for epoch J2000.0.
Columns.\ (4) and (5): estimated standard deviation of the source position error and its projected distance from cluster center.
Column.\ (6): logarithmic Poisson probability of a detection not being a source.
Columns.\ (7) -- (10): net source counts and photon flux extracted in the full (0.5--8~keV) and hard (2-8~keV) band, respectively.
Columns.\ (11) and (12): unabsorbed source luminosity in full and hard band.
The full content of this table is available online.}
\label{tab:mcat}
\end{deluxetable}

\begin{deluxetable}{@{}ll*{8}{c}rrr@{}}
\tabletypesize{\small}
\tablecolumns{8}
\linespread{1}
\tablewidth{0pc}
\tablenum{3}

\tablecaption{X-ray Source Distribution Features in M28}
\tablehead{
\colhead{Source Groups} & \colhead{R} & \colhead{$N_{\rm X}$} & \colhead{$N_{\rm B}$} & \colhead{$N_{\rm CXB}$} & \colhead{$N_{\rm G}$} & \colhead{$N_{\rm K}$} & \colhead{Ratio} \\
\colhead{(1)} & \colhead{(2)} & \colhead{(3)} & \colhead{(4)} & \colhead{(5)} & \colhead{(6)} & \colhead{(7)} & \colhead{(8)}}
\startdata
\multicolumn{8}{c}{Signal to Noise Ratio of the Distribution Dips}\\
\hline
All Sources    & 16-47   & 19 & 0.14 & 2.80 & 2.38 & 72.79 & 8.96 \\
Faint Sources  & 16-47   & 10 & 0.11 & 1.84 & 1.66 & 45.24 & 7.96 \\
Bright Sources & 13-53   & 10 & 0.03 & 1.29 & 1.09 & 34.51 & 5.81 \\
\hline
\multicolumn{8}{c}{Relative Abundance Ratio of X-ray Sources} \\
\hline
All Sources    & 0-10    & 32   & 3.6  & 0.1   & 0.1  & 24.5   & 1.44 \\
               & 10-90   & 83   & 0.7  & 10.8  & 9.2  & 137.3  & 0.46 \\
               & 0-90    & 115  & 4.3  & 10.9  & 9.3  & 161.8  & 0.61 \\
               & 90-270  & 229  & 0.0  & 65.9  & 56.1 & 52.4   & 2.04 \\
Faint Sources  & 0-9     & 16   & 2.6  & 0.1   & 0.1  & 12.5   & 1.47 \\
               & 9-70    & 41   & 0.7  & 4.4   & 3.9  & 76.9   & 0.43 \\
               & 0-70    & 57   & 3.3  & 4.5   & 4.0  & 89.4   & 0.58 \\
               & 70-200  & 118  & 0.0  & 24.1  & 21.7 & 33.6   & 2.15 \\
Bright Sources & 0-11    & 18   & 0.9  & 0.1   & 0.1  & 11.3   & 1.66 \\
               & 9-120   & 32   & 0.1  & 6.9   & 5.8  & 58.7   & 0.33 \\
               & 0-120   & 50   & 1.0  & 7.0   & 5.9  & 70.0   & 0.54 \\
               & 120-400 & 135  & 0.0  & 48.5  & 40.9 & 20.3   & 2.25 \\
\enddata
\vspace{-0.5cm}
\tablecomments{
Col.\ (1): name of the source groups defined in Figure-\ref{fig:surfd}.
Col.\ (2): the ranges (in units of arcsec) of the annulus regions.
Col.\ (3)-Col.\ (7): number of detected X-ray sources, possible count of blended sources, CXB sources, Galactic background sources and the King model predicted sources within the annulus region.
Col.\ (8): the maximum signal to noise ratio (i.e., $(N_{\rm CXB}+N_{\rm G}+N_{\rm K}-N_{\rm X}-N_{\rm B})/\sqrt{N^{2}_{\rm CXB}\sigma^{2}_{c}+N^{2}_{\rm G}\sigma^{2}_{\rm G}+N^{2}_{\rm K}\sigma^{2}_{\rm K}-N^{2}_{\rm X}\sigma^{2}_{P}(1+\beta^{2})}$) of the distribution dip and the relative abundance ratio (i.e., $(N_{\rm X}+N_{\rm B}-N_{\rm CXB}-N_{\rm G})/N_{\rm K}$) of X-ray sources in the annulus region.}
\label{tbl:spec_freq}
\end{deluxetable}

\begin{acknowledgements}
We thank the anonymous referee for the valuable comments that helped to improve our manuscript. This work is supported by China Postdoctoral Science Foundation funded project (2019TQ0288), the National Key Research and Development Program of China No. 2016YFA0400803 and No. 2017YFA0402703, the National Natural Science Foundation of China under grants 11622326, 11773015, 11873028, 11873029, and Project U1838103, U1838201 Supported by NSFC and CAS.
\end{acknowledgements}

\label{lastpage}

\end{document}